\documentclass[12pt]{article}
\usepackage[dvips]{color}
\usepackage{epsfig}
\usepackage{amsmath}
\usepackage{graphicx}

\begin{document}
\title{\bf  The Hawking temperature in the context of dark energy
for four-dimensional asymptotically AdS black holes with scalar hair}
\author{{J. Naji\thanks{Email: naji.jalil2020@gmail.com}}\\
{\small {\em Physics Department, Ilam University, Ilam, Iran}}\\
{\small {\em P.O. Box 69315-516, Ilam, Iran}}} \maketitle
\begin{abstract}
\noindent In this paper, we considered new solutions for four-dimensional asymptotically AdS black holes with scalar hair and discuss about Hawking temperature in the context of dark energy by using the tunneling method. We obtain modification of the Hawking temperature due to presence of the dark energy.
\end{abstract}
\section{Introduction}
As we know, the black holes with scalar hair are interesting solutions of Einstein's theory of gravity [1-3]. These solutions are important in the context of the no-hair theorems and are useful for application of AdS/CFT correspondence to
condense matter phenomena like superconductivity [4]. These yield to a better understanding of the behavior of matter fields from the black hole horizon to asymptotic infinity.\\
Introducing a scale in the gravity sector of the
theory through a cosmological constant is the simplest to have regular scalar field on the horizon. It is found that the hairy black
hole which minimally coupled to scalar field and a self-interaction potential in asymptotically dS
space is unstable [5, 6]. However the exact solution in asymptotically AdS space may be stable [7] where the scalar field is conformally coupled to the gravity. There are also more studies about hairy black holes in the presence of a
cosmological constant [8, 9]. Recently, new version of the hairy black hole were reported [10, 11].\\
A Reissner-Nordstrom black hole has a second-order phase
transition to a hairy black hole at the critical temperature [12].\\
Therefore, it is important to understand the behavior of a hairy black hole at
large distances. In the recent work [12] a general formalism to generate asymptotically AdS hairy black holes proposed, where the scalar field conformally coupled to
gravity with a self-interacting potential. The scalar field of this system is regular on the horizon.
Now, we would consider solutions obtained by the Ref. [12] to investigate Hawking temperature in the context of dark energy. It has been found that the Hawking temperature is modified due to the presence of dark energy [13]. Also, it is found for an emergent gravity metric having k-essence scalar fields with a Born-Infeld type lagrangian where the gravitational metric considered as Schwarzschild. In the Ref. [14] the Hawking
temperature for an emergent gravity metric in the presence
of dark energy calculated for Reissner-Nordstrom black hole metric and Kerr black
hole metric. The similar work performed by mapping the emergent gravity
metric in Schwarzschild type into a Barriola-Vilenkin black hole metric [13]. In the Ref. [14] it is found that for the case of an Reissner-Nordstrom black hole
the emergent gravity metric can be exactly mapped onto a Robinson-Trautman black hole, while For the Kerr case, the emergent metric satisfies Einstein equations for large distance.\\
In this work we would like to apply method of the Ref. [14] to possible solutions of four-dimensional asymptotically AdS black holes with scalar hair and investigate modification of the Hawking temperature due to the presence of the dark energy. This paper is organized as follows. In next section we review the emergent gravity and in section 3 we recall four-dimensional asymptotically AdS black holes with scalar hair. We present three possible solutions in sections 4-6 and discuss about the Hawking temperature. Finally in section 7 we give conclusion and summarize our results.

\section{Emergent gravity}
The emergent gravity minimal action having a
self-interacting potential $V (\phi)$ for background metric
$g_{\mu\nu}$ is given by,
\begin{equation}\label{s1}
S=\int{d^{4}x\sqrt{-g}L},
\end{equation}
where,
\begin{equation}\label{s2}
L=\frac{R}{2}-\frac{1}{2}g^{\mu\nu}\nabla_{\mu}\phi\nabla_{\nu}\phi-V(\phi),
\end{equation}
where we assumed $8\pi G=1$. The energy momentum tensor for the scalar field is,
\begin{equation}\label{s3}
T_{\mu\nu}^{(\phi)}=\nabla_{\mu}\phi\nabla_{\nu}\phi-g_{\mu\nu}[\frac{1}{2}
g^{\rho\sigma}\nabla_{\rho}\phi\nabla_{\sigma}\phi+V(\phi)].
\end{equation}
Assuming new variable,
\begin{equation}\label{s4}
X=\frac{1}{2}g^{\mu\nu}\nabla_{\mu}\phi\nabla_{\nu}\phi,
\end{equation}
we can obtain equation of motion as follow,
\begin{equation}\label{s5}
-\frac{1}{\sqrt{-g}}\frac{\delta S}{\delta\phi}=\tilde{G}^{\mu\nu}\nabla_{\mu}\nabla_{\nu}\phi+2XL_{X\phi}-L_{\phi}=0,
\end{equation}
where,
\begin{equation}\label{s6}
\tilde{G}^{\mu\nu}\equiv L_{X}g^{\mu\nu}+L_{XX}\nabla^{\mu}\phi\nabla^{\nu}\phi,
\end{equation}
with,
\begin{eqnarray}\label{s7}
L_{X}&=&\frac{dL}{dX},\nonumber\\
L_{XX}&=&\frac{d^{2}L}{dX^{2}},\nonumber\\
L_{\phi}&=&\frac{dL}{d\phi},
\end{eqnarray}
and and $\nabla_{\mu}$ is the covariant derivative with respect to the metric $g_{\mu\nu}$. Also, there is a condition as,
\begin{equation}\label{s8}
1+\frac{2XL_{XX}}{L_{X}}>0.
\end{equation}
The above formalism already applied to the The Reissner-Nordstrom black holes to study the Hawking temperature [14]. Now, we would apply them to the case of four-dimensional asymptotically AdS black holes with scalar hair given by the action (1).
\section{Four-dimensional asymptotically AdS black holes}
The general metric background of the four-dimensional asymptotically AdS black holes proposed in [12] as follow,
\begin{equation}\label{s9}
ds^{2}=-f(r)dt^{2}+\frac{dr^{2}}{f(r)}+a^{2}(r)d\sigma^{2},
\end{equation}
where $d\sigma^{2}$ is the metric of the spatial 2-section $\Sigma$, which can have positive, negative or zero curvature.\\ For the given metric it is found that,
\begin{eqnarray}\label{s10}
\tilde{G}_{00}&=&g_{00}-\dot{\phi}^{2}-V,\nonumber\\
\tilde{G}_{11}&=&g_{11}-(\phi^{\prime})^{2}-V,\nonumber\\
\tilde{G}_{22}&=&g_{22},\nonumber\\
\tilde{G}_{33}&=&g_{33}.
\end{eqnarray}
Also,
\begin{equation}\label{s11}
d\omega=dt-\left(\frac{\dot{\phi}\phi^{\prime}}{\tilde{G}_{00}}\right)dr.
\end{equation}
Therefore, the emergent gravity line element obtained as follow,
\begin{eqnarray}\label{s12}
ds_{emergent}^{2}&=&\tilde{G}_{00}dt^{2}+\tilde{G}_{11}dr^{2}-2\dot{\phi}\phi^{\prime}dtdr,\nonumber\\
&=&\tilde{G}_{00}d\omega^{2}+\left[\tilde{G}_{11}-\frac{(\dot{\phi}\phi^{\prime})^{2}}{\tilde{G}_{00}}\right]dr^{2}.
\end{eqnarray}
The emergent gravity metric will be a black hole if we have,
\begin{equation}\label{s13}
\tilde{G}_{00}=\tilde{G}_{11}^{-1}.
\end{equation}
Also, Hawking temperature given by,
\begin{equation}\label{s14}
T=\frac{r_{+}^{2}}{r_{+}-r_{-}},
\end{equation}
where $r_{+}$ and $r_{-}$ are outer and inner horizons respectively which obtained by using $f(r)=0$. We will study these for three possible solutions of the line element (9) and emergent gravity (12).
\section{The simplest MTZ solution}
The simplest solution of the line element (9) obtained by the Ref. [7] known as MTZ solution,
\begin{eqnarray}\label{s15}
f(r)&=&\frac{r(r+2\bar{\mu})}{(r+\bar{\mu})^{2}}\left(\frac{r^{2}}{l^{2}}-(1+\frac{\bar{\mu}}{r})^{2}\right),\nonumber\\
a^{2}(r)&=&\frac{r^{3}(r+2\bar{\mu})}{(r+\bar{\mu})^{2}},
\end{eqnarray}
where $\bar{\mu}\equiv\mu/8\pi$ and $l$ is the length of AdS space which related to the cosmological constant via $l^{2}=-6/\Lambda$. The mass of the solution is given by $M=2\bar{\mu}\sigma$, where $\sigma$ denotes the area of $\Sigma$.\\
The scalar field is obtained as,
\begin{equation}\label{s16}
\phi=\sqrt{6}\tanh^{-1}{\frac{\bar{\mu}}{r+\bar{\mu}}},
\end{equation}
and scalar potential is given by,
\begin{equation}\label{s17}
V(\phi)=\Lambda\sinh^{2}{\sqrt{\frac{1}{6}}\phi}.
\end{equation}
This is indeed the simplest known hairy black hole solution of a scalar field minimally coupled to the gravity which goes to zero at infinity. It is regular on the horizon given by the following radius,
\begin{equation}\label{s18}
r_{\pm}=\frac{l}{2}\left(1\pm\sqrt{1+\frac{4\bar{\mu}}{l}}\right).
\end{equation}
It tells that the black hole mass bounded $\bar{\mu}>-l/4$.\\
Hence, we can obtain the Hawking temperature as follow,
\begin{equation}\label{s19}
T=\frac{l}{4}\frac{\left(1+\sqrt{1+\frac{4\bar{\mu}}{l}}\right)^{2}}{\sqrt{1+\frac{4\bar{\mu}}{l}}}.
\end{equation}
In the case of $\bar{\mu}\gg l$ it reduces to $T=\sqrt{\bar{\mu}l}/2$.\\
The emergent gravity metric of this case given by,
\begin{eqnarray}\label{s20}
\tilde{G}_{00}&=&\frac{r(r+2\bar{\mu})}{(r+\bar{\mu})^{2}}\left(\frac{r^{2}}{l^{2}}-(1+\frac{\bar{\mu}}{r})^{2}\right)-\dot{\phi}^{2}-V,\nonumber\\
\tilde{G}_{11}&=&\frac{1}{\frac{r(r+2\bar{\mu})}{(r+\bar{\mu})^{2}}\left(\frac{r^{2}}{l^{2}}-(1+\frac{\bar{\mu}}{r})^{2}\right)}-(\phi^{\prime})^{2}-V.
\end{eqnarray}
The condition (13) suggests that,
\begin{equation}\label{s21}
\dot{\phi}^{2}=(\phi^{\prime})^{2}\left(\frac{r(r+2\bar{\mu})}{(r+\bar{\mu})^{2}}\left(\frac{r^{2}}{l^{2}}-(1+\frac{\bar{\mu}}{r})^{2}\right)\right)^{2}.
\end{equation}
We can separate $r$ and $t$-dependent sections of $\phi(r, t)$ as $R(r)+\varphi(t)$ and find,
\begin{equation}\label{s22}
\dot{\varphi}^{2}=(R^{\prime})^{2}\left(\frac{r(r+2\bar{\mu})}{(r+\bar{\mu})^{2}}\left(\frac{r^{2}}{l^{2}}-(1+\frac{\bar{\mu}}{r})^{2}\right)\right)^{2}=T,
\end{equation}
where $T$ is a constant. Therefore, we can obtain,
\begin{equation}\label{s23}
\varphi=\sqrt{T}t,
\end{equation}
and,
\begin{eqnarray}\label{s24}
R&=&\frac{2l^{2}\sqrt{T}\bar{\mu}\ln{(r+2\bar{\mu})}}{(l+4\bar{\mu})(l-4\bar{\mu})}
+\frac{(2\bar{\mu}-l)l\sqrt{T}\ln{(l\bar{\mu}+lr+r^{2})}}{4(l-4\bar{\mu})}
+\frac{(2\bar{\mu}+l)l\sqrt{T}\ln{(r^{2}-l\bar{\mu}-lr)}}{4(l+4\bar{\mu})}\nonumber\\
&+&\frac{(l-4\bar{\mu})l^{2}\sqrt{T}\tan^{-1}{(\frac{l+2r}{\sqrt{l(4\bar{\mu}-l)}})}}{2(l-4\bar{\mu})\sqrt{l(4\bar{\mu}-l)}}
-\frac{(l+4\bar{\mu})l^{2}\sqrt{T}\tanh^{-1}{(\frac{2r-l}{\sqrt{l(4\bar{\mu}+l)}})}}{2(l+4\bar{\mu})\sqrt{l(4\bar{\mu}+l)}}+C,
\end{eqnarray}
where $C$ is an integration constant. Hence we can write,
\begin{eqnarray}\label{s25}
\frac{\phi}{\sqrt{T}}&=&\frac{2l^{2}\bar{\mu}\ln{(r+2\bar{\mu})}}{(l+4\bar{\mu})(l-4\bar{\mu})}
+\frac{(2\bar{\mu}-l)l\ln{(l\bar{\mu}+lr+r^{2})}}{4(l-4\bar{\mu})}
+\frac{(2\bar{\mu}+l)l\ln{(r^{2}-l\bar{\mu}-lr)}}{4(l+4\bar{\mu})}\nonumber\\
&+&\frac{(l-4\bar{\mu})l^{2}\tan^{-1}{(\frac{l+2r}{\sqrt{l(4\bar{\mu}-l)}})}}{2(l-4\bar{\mu})\sqrt{l(4\bar{\mu}-l)}}
-\frac{(l+4\bar{\mu})l^{2}\tanh^{-1}{(\frac{2r-l}{\sqrt{l(4\bar{\mu}+l)}})}}{2(l+4\bar{\mu})\sqrt{l(4\bar{\mu}+l)}}+t+\bar{C},
\end{eqnarray}
where $\bar{C}\equiv\frac{C}{\sqrt{T}}$. The constant $T$ interpreted as kinetic energy.\\
Therefore, we can obtain emergent gravity horizon using the following equation,
\begin{equation}\label{s26}
\tilde{G}_{00}=\frac{r(r+2\bar{\mu})}{(r+\bar{\mu})^{2}}\left(\frac{r^{2}}{l^{2}}-(1+\frac{\bar{\mu}}{r})^{2}\right)-E=0,
\end{equation}
where $E=T+V$ is used. If we assume constant scalar field potential, then the total energy $E$ will be a constant. Generally we deal with a fifth order equation which their solutions give horizon radius. Assuming low mass limit and neglecting $l^{2}\bar{\mu}^{3}$ allow us to obtain the horizons for some finite values of the total energy.

\subsection{$E=-1$}
In this case we can find,
\begin{eqnarray}\label{s27}
r_{+}&=&(2\bar{\mu}l^{2})^{\frac{1}{3}},\nonumber\\
r_{-}&=&-2\bar{\mu}.
\end{eqnarray}
Therefore, Hawking temperature obtained as the follow,
\begin{equation}\label{s28}
T=\frac{(2\bar{\mu}l^{2})^{\frac{2}{3}}}{(2\bar{\mu}l^{2})^{\frac{1}{3}}+2\bar{\mu}}.
\end{equation}
Comparing with the temperature (19) we can see obvious modification due to the dark energy. Fig. 1 shows that the temperature decreased dramatically due to existence of the dark energy.

\subsection{$E=-2$}
In order to obtain explicit expression of the horizon radius of this case we should neglect also $l^{2}\bar{\mu}^{3}$ term and find,
\begin{equation}\label{s29}
r_{\pm}=-\bar{\mu}\pm\sqrt{\bar{\mu}-l^{2}},
\end{equation}
It tells that $\bar{\mu}<0$, and together previous condition we should have,
\begin{equation}\label{s30}
0>\bar{\mu}>-\frac{l}{4}.
\end{equation}
Therefore, Hawking temperature obtained as the follow,
\begin{equation}\label{s31}
T=\frac{(-\bar{\mu}+\sqrt{\bar{\mu}^{2}-l^{2}})^{2}}{2\sqrt{\bar{\mu}^{2}-l^{2}}}.
\end{equation}
Dotted line of the Fig. 1 together condition (30) shows that the hawking temperature diverges in this case.
\subsection{$E=-5$}
This case is also limited by $\bar{\mu}<0$ and yields to,
\begin{eqnarray}\label{s32}
T&=&\frac{(-45\bar{\mu}l^{2}-8\bar{\mu}^{3}+3\sqrt{192l^{6}+33l^{4}\bar{\mu}^{2}+144l^{2}\bar{\mu}^{4}})^{\frac{1}{3}}}{3}-\frac{2}{3}\bar{\mu}\nonumber\\
&-&\frac{4l^{2}-\frac{4}{3}\bar{\mu}^{2}}{(-45\bar{\mu}l^{2}-8\bar{\mu}^{3}+3\sqrt{192l^{6}+33l^{4}\bar{\mu}^{2}+144l^{2}\bar{\mu}^{4}})^{\frac{1}{3}}}.
\end{eqnarray}
Dash dotted line of the Fig. 1 shows behavior of the Hawking temperature. Having positive temperature means that negative $\bar{\mu}$ is necessary.
\subsection{$E>0$}
We can find that there is two positive real zeros of the following equation,
\begin{equation}\label{s33}
X=r^{5}+2\bar{\mu}r^{4}-l^{2}(1+E)r^{3}-2\bar{\mu}l^{2}(2+E)r^{2}-\bar{\mu}^{2}l^{2}(5+E)r-2\bar{\mu}^{3}l^{2}=0.
\end{equation}
These solutions give us the Hawking temperature and we find that it's value reduced due to the dark energy.

\begin{figure}[h!]
 \begin{center}$
 \begin{array}{cccc}
\includegraphics[width=75 mm]{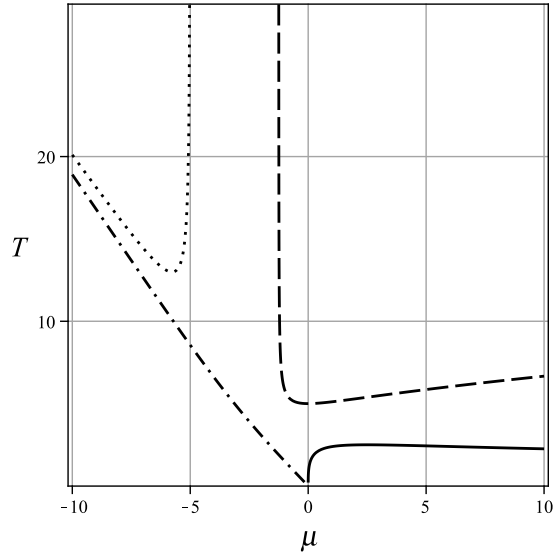}
 \end{array}$
 \end{center}
\caption{Hawking temperature in terms of $\bar{\mu}$ for $l=5$. Dashed line is corresponding to ordinary black hole, solid line is corresponding to emergent gravity solution with $E=-1$, dotted line is corresponding to emergent gravity solution with $E=-2$, dash dotted line is corresponding to emergent gravity solution with $E=-5$.}
\end{figure}
\section{The other MTZ solution}
If one wants to abandon the conformal coupling of the scalar field to the gravity deals with more complicated potential than the previous case. Such solutions obtained by [15] where new class of black hole solutions with a minimally coupled scalar field in the presence of a negative cosmological constant investigated. It is found that,
\begin{eqnarray}\label{s34}
f(r)&=&\frac{r(r+2r_{0})}{(r+r_{0})^{2}}
\left(\frac{r^{2}}{l^{2}}-\frac{gr_{0}(1+r_{0})}{l^{2}}r-1-(1-\frac{2gr_{0}^{2}}{l^{2}})\frac{r_{0}}{r}(2+\frac{r_{0}}{r})
+\frac{gr^{2}}{2l^{2}}\ln{(1+\frac{2r_{0}}{r})}\right),\nonumber\\
a^{2}(r)&=&\frac{r^{3}(r+2r_{0})}{(r+r_{0})^{2}},
\end{eqnarray}
where $r_{0}$ is a constant related to the mass.
The scalar field is obtained as,
\begin{equation}\label{s35}
\phi=\sqrt{6}\tanh^{-1}{\frac{r_{0}}{r+r_{0}}},
\end{equation}
and scalar potential is given by,
\begin{eqnarray}\label{s36}
V(\phi)&=&2\Lambda\sinh^{2}{\sqrt{\frac{1}{6}}\phi}\nonumber\\
&+&\frac{4g\Lambda}{3}\left[2\sqrt{\frac{3}{8}}\phi\cosh{\left(\sqrt{\frac{2}{3}}\phi\right)}
-\frac{9}{8}\sinh{\left(\sqrt{\frac{2}{3}}\phi\right)}
-\frac{1}{8}\sinh{\left(4\sqrt{\frac{3}{8}}\phi\right)}\right].
\end{eqnarray}
A parameter $g$ comes from conformal invariance. If $g = 0$ then the MTZ black hole solution (15) recovered (also one should change $r_{0}\rightarrow\bar{\mu}$).
On the other hand, the space becomes compact with constant negative curvature for $g \geq 2$. This solution is an asymptotically locally AdS space-time.\\
Using the dimensionless
parameter $\xi=r_{0}/r_{+}$ one can obtain the Hawking temperature as follow,
\begin{equation}\label{s37}
T=\frac{l+\xi(1+\xi)(4-g(1+2\xi+2\xi^{2}))+\frac{1}{2}g(1+2\xi)^{2}\ln{(1+2\xi)}}
{2\pi(1+2\xi)\sqrt{1+g\xi(1+\xi)(2\xi+2\xi^{2}-1)+\frac{1}{2}g\ln{(1+2\xi)}}}.
\end{equation}
\begin{eqnarray}\label{s38}
\tilde{G}_{00}&=&\frac{r(r+2r_{0})}{(r+r_{0})^{2}}
\left(\frac{r^{2}}{l^{2}}-\frac{gr_{0}(1+r_{0})}{l^{2}}r-1-(1-\frac{2gr_{0}^{2}}{l^{2}})\frac{r_{0}}{r}(2+\frac{r_{0}}{r})
+\frac{gr^{2}}{2l^{2}}\ln{(1+\frac{2r_{0}}{r})}\right)-\dot{\phi}^{2}-V,\nonumber\\
\tilde{G}_{11}&=&\frac{1}{\frac{r(r+2r_{0})}{(r+r_{0})^{2}}
\left(\frac{r^{2}}{l^{2}}-\frac{gr_{0}(1+r_{0})}{l^{2}}r-1-(1-\frac{2gr_{0}^{2}}{l^{2}})\frac{r_{0}}{r}(2+\frac{r_{0}}{r})
+\frac{gr^{2}}{2l^{2}}\ln{(1+\frac{2r_{0}}{r})}\right)}-(\phi^{\prime})^{2}-V.
\end{eqnarray}
The condition (13) suggests that,
\begin{equation}\label{s39}
\dot{\phi}^{2}=(\phi^{\prime})^{2}\left(\frac{r(r+2r_{0})}{(r+r_{0})^{2}}
\left(\frac{r^{2}}{l^{2}}-\frac{gr_{0}(1+r_{0})}{l^{2}}r-1-(1-\frac{2gr_{0}^{2}}{l^{2}})\frac{r_{0}}{r}(2+\frac{r_{0}}{r})
+\frac{gr^{2}}{2l^{2}}\ln{(1+\frac{2r_{0}}{r})}\right)\right)^{2}.
\end{equation}
The time-dependence of solution is similar to the equation (23), and for the radial part we can obtain,
\begin{equation}\label{s40}
R=\int{\frac{2\sqrt{T}(r+r_{0})^{2}l^{2}r}{(r+2r_{0})\left[2r^{4}-2l^{2}r^{2}-2gr_{0}^{2}r^{3}-4r_{0}l^{2}r+8gr_{0}^{3}r-2l^{2}r_{0}^{2}+4gr_{0}^{4}
+gr^{4}\ln{(\frac{r+2r_{0}}{r})}\right]}dr}+C,
\end{equation}
where $C$ is an integration constant. Therefore, we can obtain emergent gravity horizon using the following equation,
\begin{equation}\label{s41}
\tilde{G}_{00}=\frac{r(r+2r_{0})}{(r+r_{0})^{2}}
\left(\frac{r^{2}}{l^{2}}-\frac{gr_{0}(1+r_{0})}{l^{2}}r-1-(1-\frac{2gr_{0}^{2}}{l^{2}})\frac{r_{0}}{r}(2+\frac{r_{0}}{r})
+\frac{gr^{2}}{2l^{2}}\ln{(1+\frac{2r_{0}}{r})}\right)-E=0,
\end{equation}
where $E=T+V$ is used as before.\\
Neglecting logarithmic part and assuming $2r_{0}^{2}(2gr_{0}^{2}-l^{2})\ll1$ allow us to obtain horizon radius and then Hawking temperature. Generally we should solve the following equation,
\begin{equation}\label{s42}
r^{4}+Ar^{3}+Br^{2}+Cr+D=0,
\end{equation}
where,
\begin{eqnarray}\label{s43}
A&=&r_{0}(2-g-gr_{0}),\nonumber\\
B&=&-(l^{2}(1+E)+2gr_{0}(1+r_{0})),\nonumber\\
C&=&2r_{0}(2gr_{0}^{2}-l^{2}(2+E)),\nonumber\\
D&=&r_{0}^{2}(10gr_{0}^{2}-l^{2}(5+E)),
\end{eqnarray}
Which recovers results of the previous section for $g=0$ limit.\\
A possible solution of the equation (42) will obtained under assumptions $B=0$, $C=-A$ and $D=-A^{2}$. They mean that,
\begin{eqnarray}\label{s44}
g&=&\frac{2(l^{2}-1)}{8r_{0}^{2}+3r_{0}-1},\nonumber\\
E&=&\frac{4r_{0}^{2}(1-3l^{2})+r_{0}(4-7l^{2})+l^{2}}{l^{2}(8r_{0}^{2}+3r_{0}-1)},
\end{eqnarray}
and $r_{0}$ is the root of the following equation,
\begin{equation}\label{s45}
16r_{0}^{4}+38r_{0}^{3}+(19-l^{2})r_{0}^{2}+(1-2l^{2})r-l^{2}=0.
\end{equation}
There are some possible values such as $l=\sqrt{1/2}$, $r_{0}\approx0.14$, $g=2.47$, and $E=-2.62$, or $l=1$, $r_{0}\simeq0.21269$, $g=0$, and $E\approx-3$.\\
Under above assumptions we can obtain solution of the equation (42) as follow,
\begin{eqnarray}\label{s46}
r_{+}&=&(r_{0}(2-g(1+r_{0})))^{\frac{1}{3}},\nonumber\\
r_{-}&=&-r_{0}(2-g(1+r_{0})).
\end{eqnarray}
This is indeed extension version of the previous case with $E=-1$ to including $g$.
Therefore, Hawking temperature obtained as the follow,
\begin{equation}\label{s47}
T=\frac{(r_{0}(2-g(1+r_{0})))^{\frac{2}{3}}}{(r_{0}(2-g(1+r_{0})))^{\frac{1}{3}}+r_{0}(2-g(1+r_{0}))}.
\end{equation}

\begin{figure}[h!]
 \begin{center}$
 \begin{array}{cccc}
\includegraphics[width=75 mm]{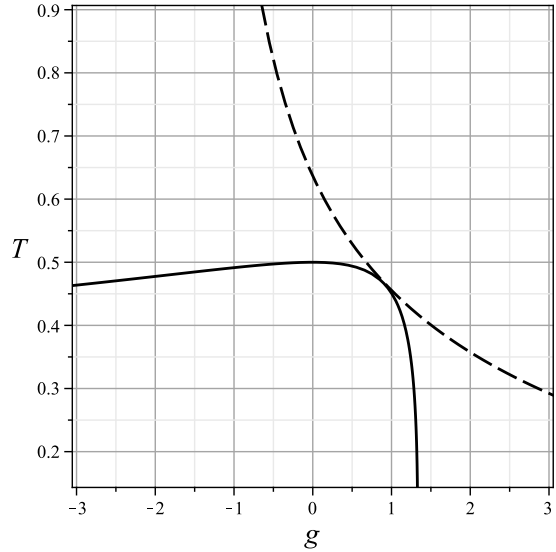}
 \end{array}$
 \end{center}
\caption{Hawking temperature in terms of $g$ for $l=5$, $\xi=r_{0}=0.5$. Dashed line is corresponding to ordinary black hole, solid line is corresponding to emergent gravity solution.}
\end{figure}

\section{Asymptotically AdS solution with scalar hair}
Interesting solution of the line element (9) recently obtained by [12]. It is found that,
\begin{eqnarray}\label{s48}
f(r)&=&k+rF(r+\nu)+\frac{G}{\nu^{3}}\left(2r(r+\nu)\ln{(\frac{r+\nu}{r})}-\nu(\nu+2r)\right),\nonumber\\
a(r)&=&\alpha r^{\frac{1}{2}(1+\sqrt{1-2b^{2}})}(r+\nu)^{\frac{1}{2}(1-\sqrt{1-2b^{2}})}+\frac{\beta r^{\frac{1}{2}(1-\sqrt{1-2b^{2}})}(r+\nu)^{\frac{1}{2}(1+\sqrt{1-2b^{2}})}}{\nu\sqrt{1-2b^{2}}},
\end{eqnarray}
where $k = -1, 0, 1$ corresponding to space curvature and $F$, $G$ are constants being proportional to the cosmological constant and the mass respectively. Also, $\alpha$ and $\beta$ are arbitrary integration constants. It is possible to choose $\alpha=1$, $\beta=0$ and $b=1/\sqrt{2}$ to find,
\begin{equation}\label{s49}
a^{2}(r)=r(r+\nu),
\end{equation}
where $\nu$ is free parameter. Indeed the $f(r)$ given by the relation (48) obtained by using the above fixed parameters.\\
The scalar field is obtained as,
\begin{equation}\label{s50}
\phi=\frac{1}{\sqrt{2}}\ln{(1+\frac{\nu}{r})},
\end{equation}
and scalar potential is given by,
\begin{equation}\label{s51}
V(\phi)=-\frac{1}{2\nu^{3}r(r+\nu)}\left[-6\nu G(2r+\nu)+(6r^{2}+6r\nu+\nu^{2})\left(\nu^{3}F+2G\ln{(\frac{r+\nu}{r})}\right)\right].
\end{equation}
It has been shown that the above solution reduced to the Schwarzschild AdS solution for $k=1$ and topological AdS black hole solution for $k=-1$ at $\phi=0$ limit [12].\\
Assuming large value for the parameter $\nu$ comparing to the coordinate $r$ allows us to consider logarithmic part as a constant. In that case one can obtain horizon radius as follow,
\begin{equation}\label{s52}
r_{\pm}=\frac{2G(1-\ln{\nu})-F\nu^{3}\pm\sqrt{\delta}}{2(F\nu^{3}+2G\ln{\nu})}\nu,
\end{equation}
where we defined,
\begin{equation}\label{s53}
\delta\equiv F^{2}\nu^{6}+4G^{2}-4Fk\nu^{4}+4G(F\nu^{3}-2k\nu+G\ln{\nu})\ln{\nu}.
\end{equation}
Hence, the Hawking temperature reads as,
\begin{equation}\label{s54}
T=\frac{\nu[F\nu^{3}-2G(-1+\ln{\nu})-\sqrt{\delta}]^{2}}{4(F\nu^{3}+2G\ln{\nu})\sqrt{\delta}}.
\end{equation}
We can investigate the temperature for possible values $F=-1, 0, +1$ and $k=-1, 0, +1$ to find $F<0$ is forbidden and yields to negative temperature. Also, the case of $F=0$ together $k=1$ yield to negative temperature. All other cases are valid and represented in the Fig. 3. We can see that $F=k=0$ is valid only for $G>0$. In the case of $F=0$ and $k=-1$ the Hawking temperature is decreasing function of $G$.\\

\begin{figure}[h!]
 \begin{center}$
 \begin{array}{cccc}
\includegraphics[width=50 mm]{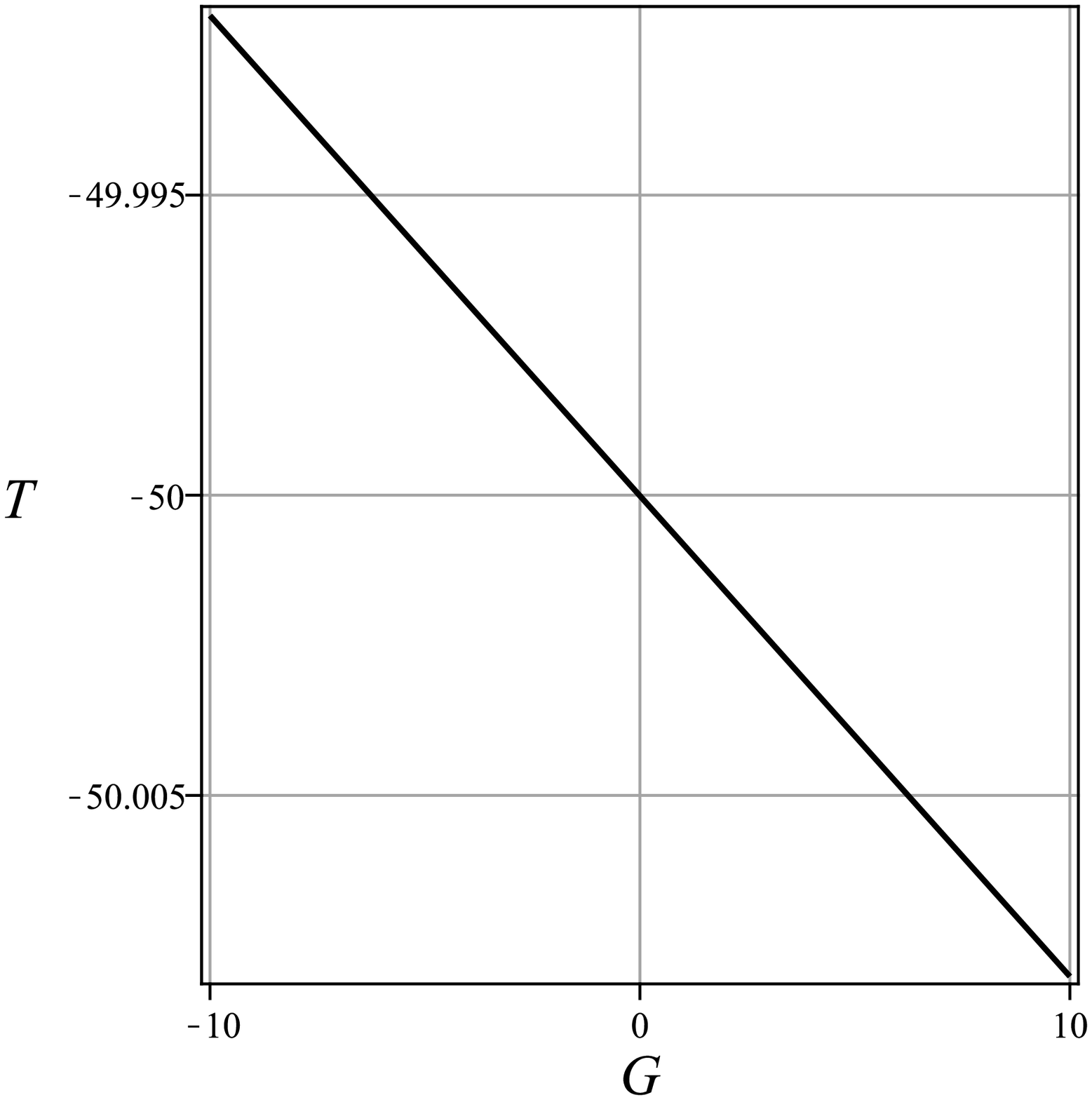}\includegraphics[width=50 mm]{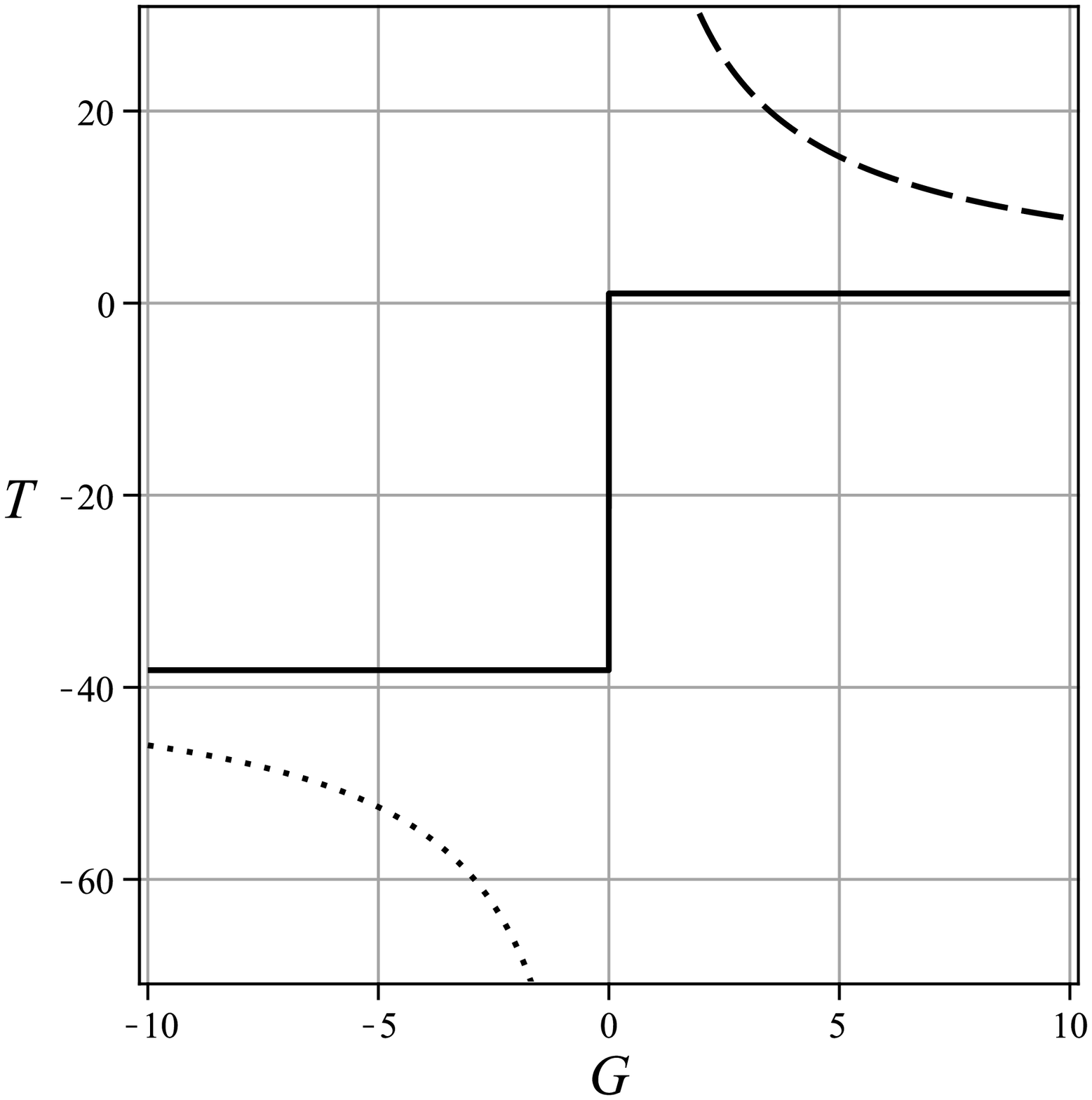}\includegraphics[width=50 mm]{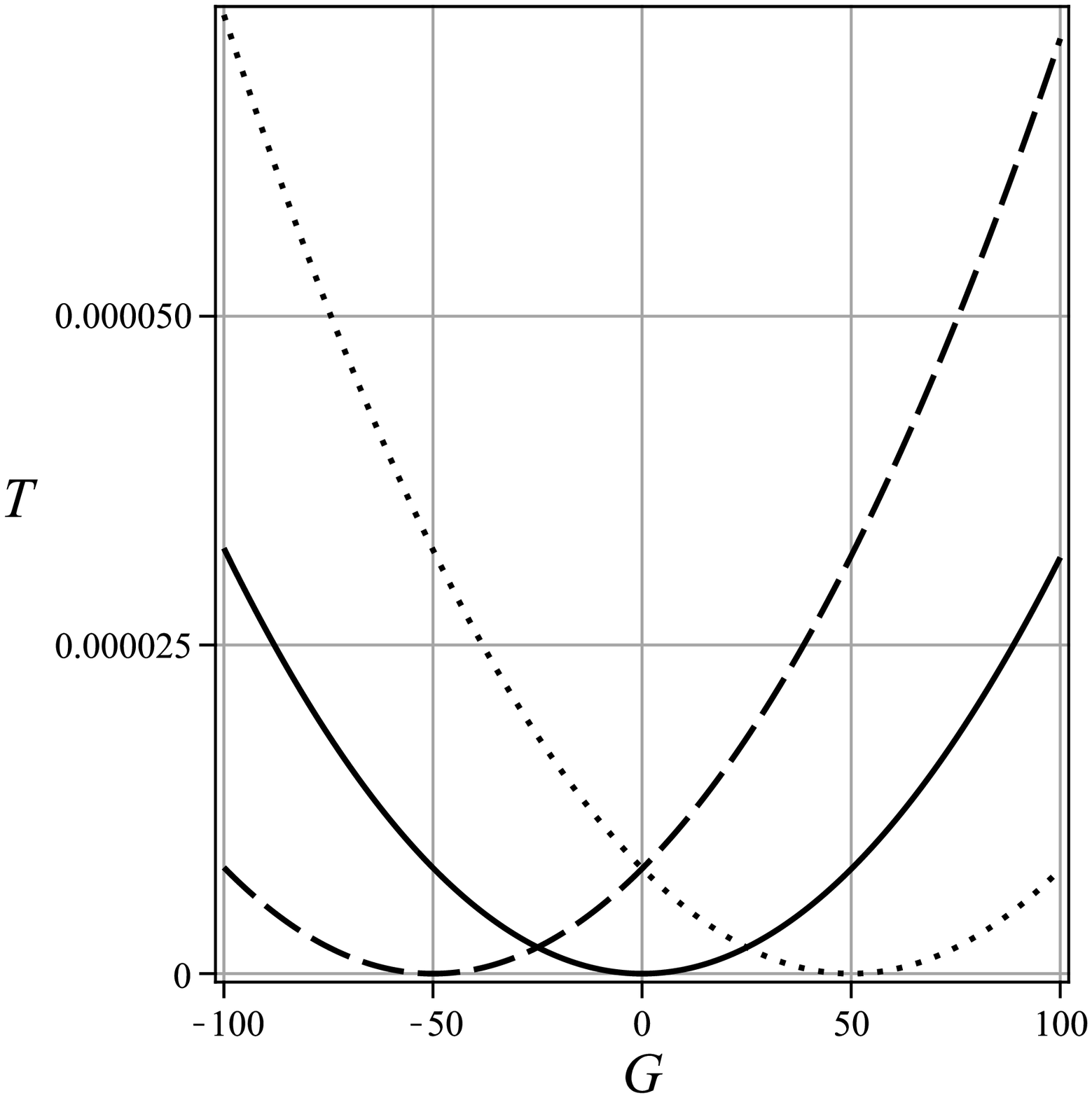}
 \end{array}$
 \end{center}
\caption{Hawking temperature in terms of $G$ for $\nu=50$. Left: $F=-1$, $k=-1$ (dashed line), $k=0$ (solid line), $k=+1$ (dotted line) (in this case all curves are coincide). Middle: $F=0$, $k=-1$ (dashed line), $k=0$ (solid line), $k=+1$ (dotted line). Right: $F=+1$, $k=-1$ (dashed line), $k=0$ (solid line), $k=+1$ (dotted line).}
\end{figure}

The emergent gravity metric of this case given by,
\begin{eqnarray}\label{s55}
\tilde{G}_{00}&=&k+rF(r+\nu)+\frac{G}{\nu^{3}}\left(2r(r+\nu)\ln{(\frac{r+\nu}{r})}-\nu(\nu+2r)\right)-\dot{\phi}^{2}-V,\nonumber\\
\tilde{G}_{11}&=&\frac{1}{k+rF(r+\nu)+\frac{G}{\nu^{3}}\left(2r(r+\nu)\ln{(\frac{r+\nu}{r})}-\nu(\nu+2r)\right)}-(\phi^{\prime})^{2}-V.
\end{eqnarray}
The condition (13) suggests that,
\begin{equation}\label{s56}
\dot{\phi}^{2}=(\phi^{\prime})^{2}\left(k+rF(r+\nu)+\frac{G}{\nu^{3}}\left(2r(r+\nu)\ln{(\frac{r+\nu}{r})}-\nu(\nu+2r)\right)\right)^{2}.
\end{equation}
Similar to the previous section we can rewrite $\phi(t, r)=\varphi(t)+R(r)$ and find,
\begin{equation}\label{s57}
\dot{\varphi}^{2}=(R^{\prime})^{2}\left(k+rF(r+\nu)+\frac{G}{\nu^{3}}\left(2r(r+\nu)\ln{(\frac{r+\nu}{r})}-\nu(\nu+2r)\right)\right)^{2}=T,
\end{equation}
where $T$ is a constant as before and interpreted as kinetic energy. In order to obtain the function $R$ assume large value of $\nu$ to fix logarithmic part. In that case one can obtain,
\begin{equation}\label{s58}
\phi=\frac{2\sqrt{T}\nu^{3}}{\sqrt{\delta}}\tan^{-1}{\left(\frac{2(F\nu^{3}+2G\ln{\nu})r+F\nu^{4}-2G\nu+2G\nu\ln{\nu}}{\sqrt{\delta}}\right)}+\sqrt{T}t+C,
\end{equation}
where $C$ is an integration constant. Therefore, we can obtain emergent gravity horizon using the following equation,
\begin{equation}\label{s59}
\tilde{G}_{00}=k+rF(r+\nu)+\frac{G}{\nu^{3}}\left(2r(r+\nu)\ln{(\frac{r+\nu}{r})}-\nu(\nu+2r)\right)-E=0,
\end{equation}
where $E=T+V$ is used as before. In that case the Hawking temperature (54) modified as follow,
\begin{equation}\label{s60}
T=\frac{\nu[F\nu^{3}-2G(-1+\ln{\nu})-\sqrt{\delta+4E\nu(F\nu^{3}+2G\ln{\nu})}]^{2}}{4(F\nu^{3}+2G\ln{\nu})\sqrt{\delta+4E\nu(F\nu^{3}+2G\ln{\nu})}}.
\end{equation}
In the Fig. 4 we can see effect of the $E$ on the temperature. Significant results is that the system with $F=1$, $k=0$, $E=-1$ is corresponding to the system with $F=1$, $k=1$, $E=0$, and the system with $F=1$, $k=0$, $E=+1$ is corresponding to the system with $F=1$, $k=-1$, $E=0$. It means that the effect of the dark energy is changing curvature constant of the model.\\
We should note that this significant feature obtained for special cases of $F=0, \pm1$, $k=0, \pm1$ and $E=0, \pm1$. So, other values may yield to different results.\\

\begin{figure}[h!]
 \begin{center}$
 \begin{array}{cccc}
\includegraphics[width=50 mm]{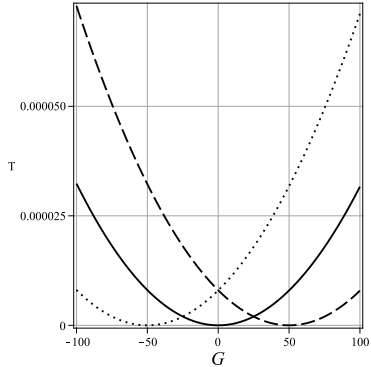}
 \end{array}$
 \end{center}
\caption{Hawking temperature in terms of $G$ for $\nu=50$, $F=+1$ and $k=0$. $E=-1$ (dashed line), $E=0$ (solid line), $E=+1$ (dotted line).}
\end{figure}
\section{Conclusion}
In this work, we studied the effect of dark energy on the Hawking temperature for some classes of the four-dimensional
asymptotically AdS black holes with scalar hair. The same study already performed for the Reissner-Nordstrom and Kerr black holes [14] and found that the Hawking temperature of the resulting emergent gravity will zero and the black hole does not radiate. This is not happen at all for other black holes such as the four-dimensional
asymptotically AdS black holes with scalar hair which we considered in this paper. However we can obtain always some condition where the Hawking temperature vanishes. For example, in the simplest MTZ solution which discussed in section 4 we have shown that for the case of $E=-2$ that some negative $\mu$ yields to zero temperature and the resulting emergent gravity black hole does not radiate. Similar condition may be found in the other MTZ solution which discussed in section 5. We found that some values of $g$ in the interval $1<g<1.5$ may yield to the zero temperature. Also, we found that some values of $g$ in the interval $0.5<g<1$ may canceled the effect of dark energy and both temperature will be the same. Similar situation exist for the asymptotically AdS solution with scalar hair which discussed in section 6. In general, we found that the presence of the dark energy may increase or decrease value of the Hawking temperature depending on the curvature constant, cosmological constant and the black hole mass.

\end{document}